\documentclass{elsart}

\usepackage{epsfig}
\usepackage{amssymb}

\def\Journal#1#2#3#4{{#1} #2 (#3) #4}
\def\NIMA{Nucl. Instr. Meth. A}
\def\NPBProc{Nucl. Phys. B (Proc. Suppl.)}
\def\PRD{Phys. Rev. D}
\def\PLB{Phys. Lett. B}
\def\ApJ{Astrophys. J}
\def\AstJ{Astron. J.}
\def\AstroPar{Astropart. Phys.}

\begin{document}

\begin{frontmatter}

\title{First results with the ORPHEUS dark matter detector}

\author{K.~Borer, G.~Czapek, F.~Hasenbalg, M.~Hauser, S.~Janos, }
\author{P.~Loaiza, U.~Moser,} 
\author{K.~Pretzl\corauthref{cor1}},
\author{C.~Sch\"utz, P.~Wicht,}
\author{and S.~W\"uthrich}
\address{Laboratory for High Energy Physics, University of Bern, 
        Sidlerstrasse 5, CH 3012 Bern, Switzerland}
\corauth[cor1]{Corresponding author. Tel.: +41--31--631--8566;
fax: +41--31--631--4487.\\
{\em E--mail address:} klaus.pretzl@lhep.unibe.ch (K.~Pretzl).}

\title{}

\author{}

\address{}

\begin{abstract}
The ORPHEUS dark matter detector is operating at our underground laboratory
in Bern (70~m.w.e.). The detector relies on measuring the magnetic flux
variation produced by weakly interacting massive particles (WIMPs) as they 
heat micron--sized superheated superconducting tin granules (SSG) and induce
superconducting--to--normal phase transitions. In an initial phase, 0.45~kg
of tin granules in a segmented detector volume have been used. In this paper
a general description of the experimental set-up, overall performance of the
detector, and first results are presented.
\end{abstract}

\end{frontmatter}

\section{Introduction}

According to the latest cosmic microwave background measurements \cite{Spergel}
combined with other cosmological observations 
\cite{SNIa,SNIa2,2dFGRS,SDSS,lensing}, the matter density of the 
universe comprises 27\% (for a Hubble constant $H_{\rm o} = 
71$~km~s$^{-1}$~Mpc$^{-1}$) of the critical density  while the remaining 73\%
is in the form of a cosmological constant of an unknown nature. With only 
$\approx$~4\% of the matter density in the form of baryons, the rest should 
be of some exotic non--baryonic nature. This non--luminous and 
non--relativistic matter, known as cold dark matter, constitutes an 
essential ingredient in theories of structure formation in the universe. 
From the class of candidates for cold dark matter generically known as 
weakly interacting massive particles (WIMPs), the lightest supersymmetric 
particle, the neutralino~\cite{Jungman}, seems to be the most promising. 
The upper bound for the mass of the neutralino in the constrained minimal 
supersymmetric extension of the Standard Model is 500~GeV~\cite{Ellis}, 
its lower bound of 46~GeV comes from $e^+ e^-$ collider experiments~\cite{LEP}.

The ORPHEUS project is a direct search experiment looking for cold dark matter
aiming at detecting the small energy deposited by a neutralino after being 
elastically scattered off a target nucleus~\cite{Ablanalp}. The detector consists
of a homogeneous mixture of superheated superconducting tin granules (SSG)
embedded in a dielectric filling material and exposed to an external magnetic
field. The type~I superconducting granules are kept slightly below
the boundary of their superconducting--to--normal phase transition in a
metastable state. The recoil energy released by a particle interacting with
a granule causes a temperature increase inversely proportional to the specific
heat of the granule. Due to the small value of the specific heat at
low temperatures, the deposited energy is enough to make a granule 
normal--conducting. A 1~keV energy deposition in a granule of 35~$\mu$m 
diameter at a bath temperature of 110~mK increases the temperature of the 
granule by typically 100~mK. The induced flux change ({\em flip}) due to the 
disappearance of the Meissner--Ochsenfeld effect is measured by an RLC circuit.
A general description of SSG detectors is given in 
Refs.~\cite{Drukier,Pretzl,Pretzl2}.

This paper is organised as follows: section 2 describes the experimental 
setup, the shielding, and the data acquisition system; section 3 describes
the data taking and its reduction procedure as well as the detector response
to muons, section 4 outlines the main results of this experiment and
presents the inferred exclusion plot. Finally, section 5 discusses possible 
improvements of the detector capabilities and outlook. An appendix briefly 
describes how the expected rates in our detector are computed.

\section{Experimental setup}

\begin{figure}[h]
\centering
\epsfysize=9.3truecm
\epsffile{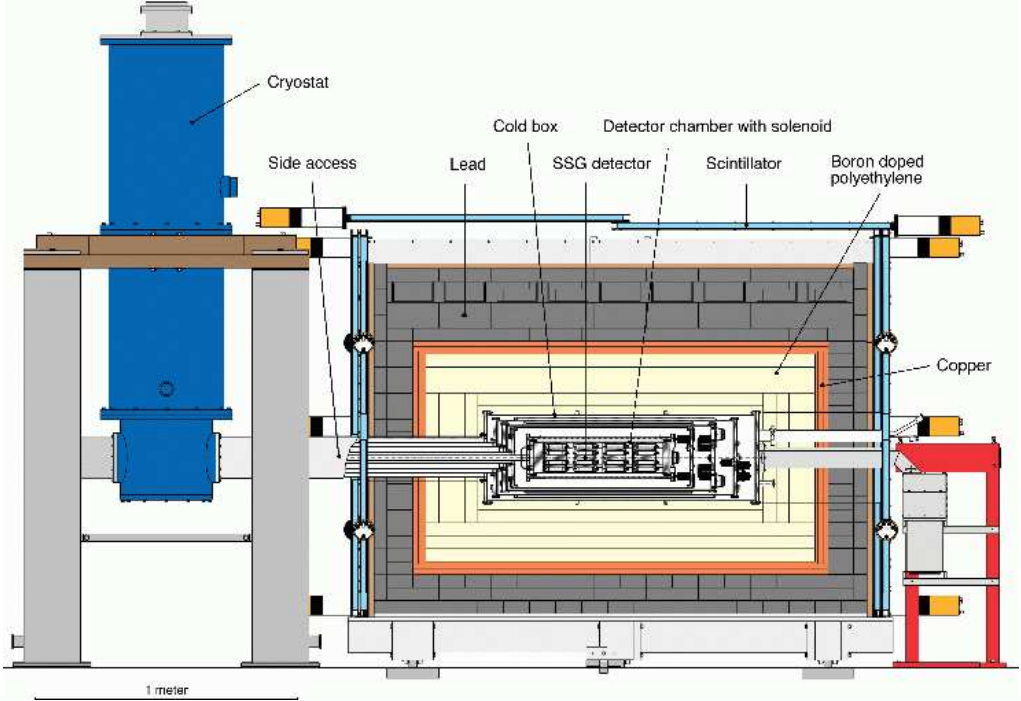}
\caption{ORPHEUS experimental setup. \label{setup}}
\end{figure}

ORPHEUS is located in the underground facility of the University of Bern
(70~m.w.e.). A scheme of the experimental setup is shown in Fig.~\ref{setup}.
It consists of the detector chamber, the electronic readout, the cryogenic 
system, the passive shielding, and the muon veto. Each part is described in 
some detail below.

\subsection{SSG detector and detector chamber}

Tin granules, of 5N purity, with two different diameter ranges are selected as
sensitive material. The granules are produced by the Technical University of
Clausthal using a fine powder gas atomisation technique and sieved in Bern to
the desired diameter ranges. Out of 50~kg of granules of different sizes only
$\approx$~1~kg remain after sieving. A sample of granules of both diameter
ranges was measured with an optical scanning microscope. The sample of 
large granules was found to have a nearly Gaussian distribution with a
mean diameter $\hat{\phi} = 36.6$~$\mu$m and a standard deviation of $\sigma$
= 2.2~$\mu$m. The sample of small granules also exhibited a Gaussian 
distribution but with a mean diameter $\hat{\phi} = 27.7$~$\mu$m and $\sigma$
= 4.1~$\mu$m. To avoid magnetic interactions among the granules, we use Teflon
powder as a filling material at a volume filling factor for tin of 10\%. This
corresponds roughly to a mean distance between the granule centres of about 
2 granule diameters.

\begin{figure}[h]
\centering
\epsfxsize=13.9truecm
\epsffile{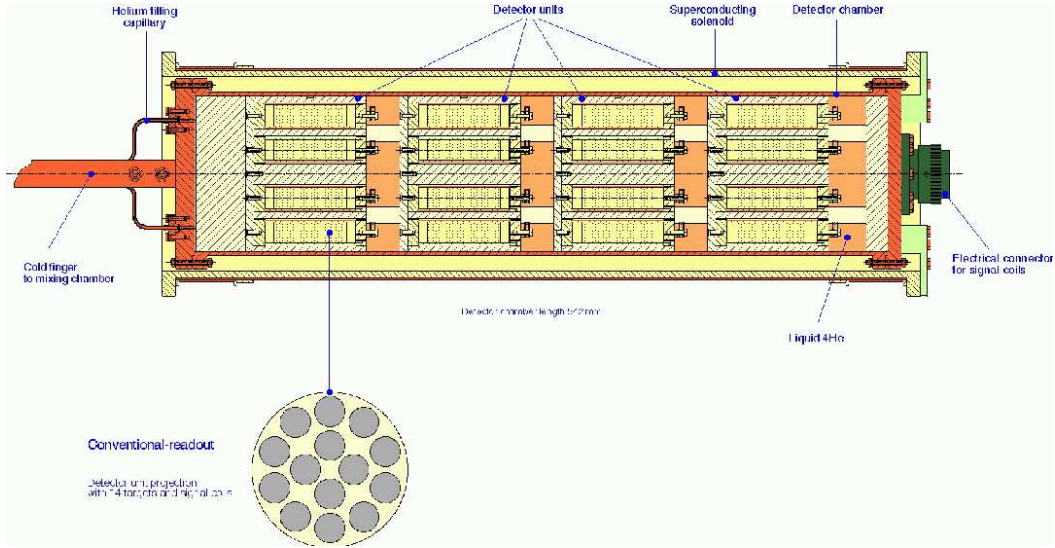}
\caption{ORPHEUS detector chamber. \label{detch}}
\end{figure}

The detector chamber consists of 56 pick--up coils filled with sensitive
material. The coils, 1.8~cm in diameter (1.6~cm inner diameter), 6.8~cm long
and with roughly 1500 windings, were filled half of them with large granules 
and the other half with small ones. A filled coil contains $\approx$ 8~g of 
tin. A total mass of 207~g was measured for the large granules (excluding 2
dead channels) and 215~g for the small ones (excluding 1 dead channel). The
dead channels were due to wiring problems at the pick--up coil and register
no output signal. Inside the chamber 4 detector modules made of Delrin, serve
as a holder for the pick--up coils with 14 coils in each module. A test coil
with 10 windings around every detector module generates a test signal. 
Fig.~\ref{detch} shows the detector chamber with the four modules and the
superconducting solenoid surrounding it. This configuration provides a highly
segmented detector which is very useful to locate hot spots and to reject 
spurious events. 

\subsection{Electronic readout}

ORPHEUS uses a damped resonance parallel RLC circuit to generate a voltage
pulse when a flip occurs. The coils have a typical inductance of $L$ = 10~mH,
cooled shunt resistors of $R$ = 10~k$\Omega$, and capacitances C $\approx $ 
1~nF determined by the combination of cable and input capacitance of the low 
noise preamplifier. The resonance frequency of the circuit is $f \approx 
50$~kHz. 

\begin{figure}[h]
\centering
\epsfysize=9.3truecm
\epsffile{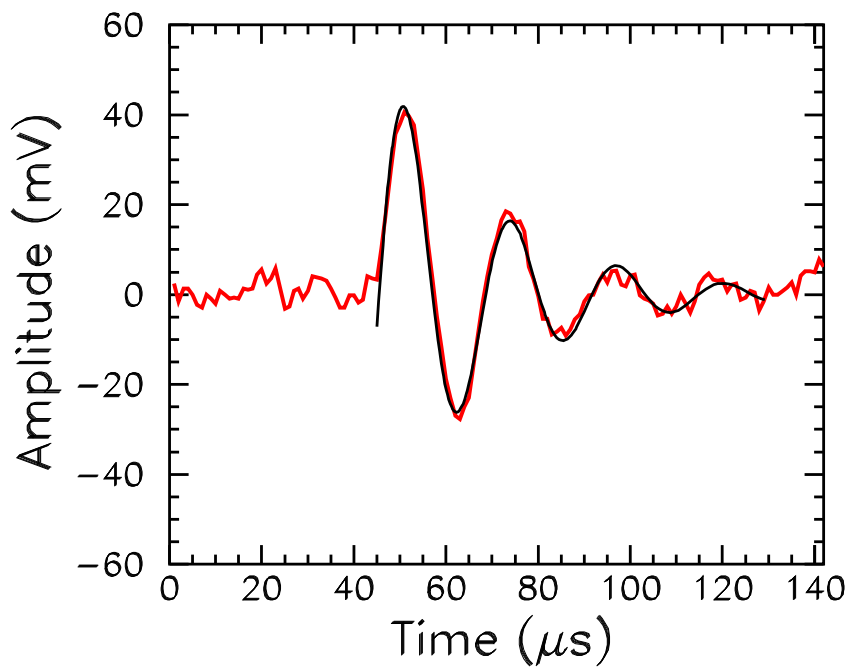}
\caption{A typical granule flip signal ($\phi \approx$ 36 $\mu$m) with
its fit. \label{flip}}
\end{figure}

The change in magnetic flux $\Delta \Phi$ caused by a flipping granule of 
radius $r$, at applied magnetic field $B_a$, in the centre of a coil of 
length $l$, radius $R$, and $n$ windings is, 
\begin{equation}
   \Delta \Phi\; =\; 2 \pi B_a n \, \frac{r^3}{\sqrt{4 R^2 \, +\, l^2}}.
\end{equation}
If the flipping time $\tau$ is small compared to the characteristic period of
the readout circuit ($\tau << 2 \pi \sqrt{LC}$), the flux variation induces a
voltage pulse in the pick--up coil of the form,
\begin{equation}
   V(t)\; =\; \frac{\Delta \Phi}{\omega LC} \, e^{-t/2RC}\,\sin(\omega t)
              \hspace{1.5cm} \omega^2 \;=\; 1/LC - 1/(2RC)^2.
\end{equation}
This signal of a few $\mu$V is amplified by a factor $10^4$ with low
noise JFET amplifiers at room temperature. Only 80\% of the coil volume is 
filled with granules in order to get uniform signals within the sensitive 
volume. A typical granule flip signal ($\phi \approx$ 36~$\mu$m) fitted to a 
function  of the form (2) can be seen in Fig.~\ref{flip}. The signal to noise 
ratio is between 10 and 20, depending on the granule size. A 
detailed description of the readout concept and noise estimations is given in 
Ref. \cite{Borer}.
 
\subsection{Cryogenic system}

The detector chamber is kept at a base temperature of 115~mK with a 
temperature stability of $\pm$~5~mK using a modified Oxford $^3$He/$^4$He 
dilution refrigerator of 300~$\mu$W cooling power. The cold box consists of
concentric copper thermal shields at temperatures of 1.7, 4.5, and 77~K
around the detector. Because of the limited vertical space available in the
Bern underground facility, a side access of 1.4~m was necessary to connect 
the mixing chamber of the dilution unit with the detector chamber. All thermal
shields of the dilution refrigerator are brought in thermal contact with 
their corresponding thermal shields on the cold box using flexible bellows. 
The ``L--shaped'' configuration poses stringent cooling demands on the 
dilution refrigerator which were solved by installing two additional cooling 
jackets, one with liquid N$_2$ and one with liquid He, in contact with the 
thermal shields of the cold box through which liquid refrigerants are kept 
flowing continuously. 

The He consumption is typically of 1~l~h$^{-1}$ for the dilution refrigerator
and 2.6~l~h$^{-1}$ for the additional cooling. The total time required to
reach the base temperature is approximately 2 weeks. The outer vacuum can 
and the inner vacuum can operate typically at 10$^{-6}$~mbar.

\subsection{Shielding and muon veto}

The passive shielding reduces the environmental radioactive flux while the
active shielding efficiently vetoes the flux produced by muons from cosmic 
rays. The passive shielding consists of 15~cm of lead, 4~cm of oxygen free 
high conductivity copper (OFHC), and 18~cm of boron--doped (5\%) polyethylene
to moderate and capture environmental neutrons. The detector chamber is made 
of electroformed copper to guarantee a low radioactivity level. The shielding, 
resting on wheels, can be opened in two halves to access the detector.

Since ORPHEUS is operated at a shallow depth, active shielding is necessary. 
Muons are registered by 2~cm thick plastic scintillators (NE 102A) surrounding
the passive shielding and analysed off--line to reject signals from the 
detector chamber in coincidence with the muons within a time window of 15 
$\mu$s. The plastic scintillators are coupled through wavelengthshifter bars 
to one or two photomultiplier tubes. The total veto--trigger rate is 
approximately 5~kHz. At our depth, the expected flux of muon--induced neutrons
in 15~cm of lead is 2~x~10$^{-4}$ n cm$^{-2}$~s$^{-1}$ \cite{Da Silva}. We 
estimate that the muon veto reduces this flux by more than a factor 100. 

\begin{table}[h]
\begin{center}
\begin{tabular}{c c c c c c c} \hline \hline
Probe & \multicolumn{6}{c}{Activities (mBq~kg$^{-1}$)} \\
      & $^{238}$U & $^{232}$Th & $^{210}$Pb & $^{137}$Cs & $^{60}$Co & $^{40}$K
\\
\hline 
Sn granules   &   $<$7      &   $<$8      &   $<$1800    &   $<$7     &   $<$4    
&    $<$72     \\
Teflon powder   &  $<$9      &   $<$11     &   $<$70      &   $<$10    &   $<$9    
&    $<$108    \\
nylon screws & $<$14     &   $<$14     &   $<$70      &   $<$13    &   $<$7    
&    $<$139    \\
solenoid wire & $<$3      &   $<$4      &   $<$550     &   $<$2     &   $<$1    
&    $<$25     \\      
Delrin    & $<$10     &   $<$12     &   $<$91      & 21 $\pm$ 5 &   $<$9     &   
$<$132    \\
OFHC Cu       & $<$82     &   $<$136    &   $<1.2\times10^4$  &   $<$86    &  
$<$65    &    $<$1200   \\
\vspace{-0.3cm}
connector  & 314$\pm$5 & 295$\pm$8 & 600$\pm$15 & 18$\pm$2 & 24$\pm$2 &
4600$\pm$90 \\
(280 g) & & & & & & \\
Pb shielding      &           &             & $(2\pm0.4)\times10^5$ &  &            
&          \\
\hline \hline \\
\end{tabular} 
\end{center}
\caption{Sample activities in mBq~kg$^{-1}$ at 95\% CL. \label{activities}}
\end{table}

A series of radiopurity measurements of the detector components were
performed with a low--background germanium detector in the underground 
facility of the Gotthard tunnel (3000~m.w.e). Several probes were measured 
for their activity: high purity Sn granules (used in ORPHEUS), Teflon powder 
(polytetrafluorethylene, PTFE), nylon screws, solenoid wire, Delrin 
(polyacetal polyoxymethylene, POM), OFHC Cu, a feedthrough connector, and
several Pb samples. A summary of the values obtained is shown in Table 
\ref{activities}. Not all the samples have the same sensitivities due to 
differences in the masses of the probes and in measuring times. 

A major concern is the activity of the ORPHEUS lead shield. Several probes 
were measured for their $^{210}$Pb content by $\alpha$ and $\gamma$ 
spectroscopy. It was found that the ORPHEUS lead has on average an activity 
of $\approx$~200~Bq~kg$^{-1}$.

\subsection{Data acquisition}

\begin{figure}[h]
\centering
\epsfysize=13.truecm
\epsffile{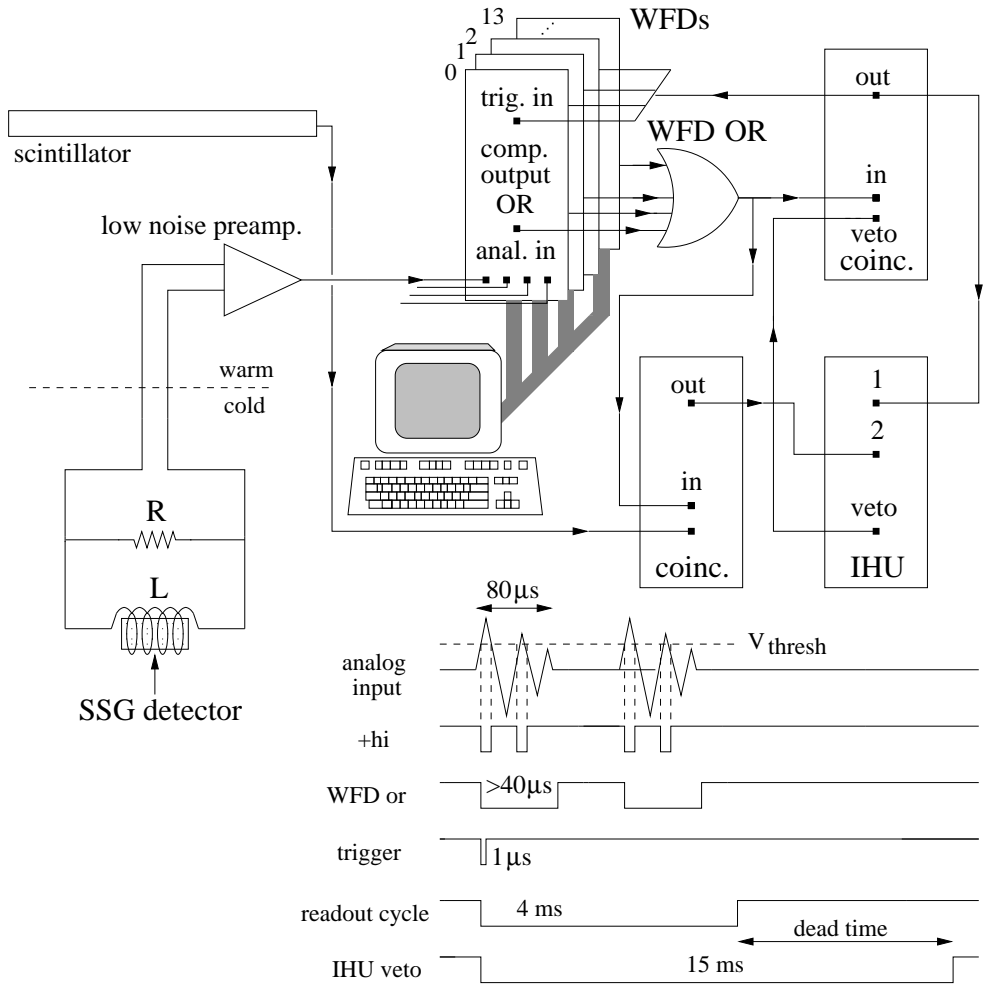}
\vspace{0.5cm}
\caption{Readout circuit, data acquisition and timing scheme. \label{daq}}
\end{figure}

To perform off--line pulse shape and background rejection analysis,
the phase transition signals are further amplified and digitised 
in custom--made CAMAC modules or waveform digitisers (WFD). Each WFD has 4 
independent channels with differential amplifiers, analog--to--digital 
converters, memory buffers and voltage comparators providing a trigger signal.
Typically 150 points per pulse were digitised at a sampling rate of 1 
mega sample per second. A scheme of the readout circuit and data acquisition
system can be seen in Fig.~\ref{daq}.

The OR of all 56 comparators is used as an event trigger (see Fig.~\ref{daq}).
After each event trigger, the interrupt handling unit (IHU) starts the 
readout of all WFD modules and vetoes further triggers during the readout 
time of 15~ms. The WFDs continue to record the signals during 4.1~ms after
the event trigger (memory buffer of 4096 words at 1~$\mu$s/word), which
results in an effective dead time of 11~ms per event. Therefore, at a typical
trigger rate in the range of 1~Hz to 10~Hz the dead time was 1\% to 11\%.

Muon events are rejected off--line if an event is coincident with a
plastic scintillator signal within 15~$\mu$s. They are flagged by means of
a signal present at the input 2 of the interrupt handling unit (see 
Fig.~\ref{daq}). Accidental coincidences are also registered measuring the 
coincidences of the WFD OR with the scintillator signals delayed by 40~$\mu$s. 

\section{Data taking}

We were able to take data continuously during two months in 2002 at a
temperature of 170~mK \cite{taup01,idm02,ltd10}, and during 3 weeks in 2003 
at 115~mK. An upgraded acquisition system, as described in the previous 
section, was implemented in the 2003 run.

\begin{figure}[h]
\centering
\epsfysize=9.3truecm
\epsffile{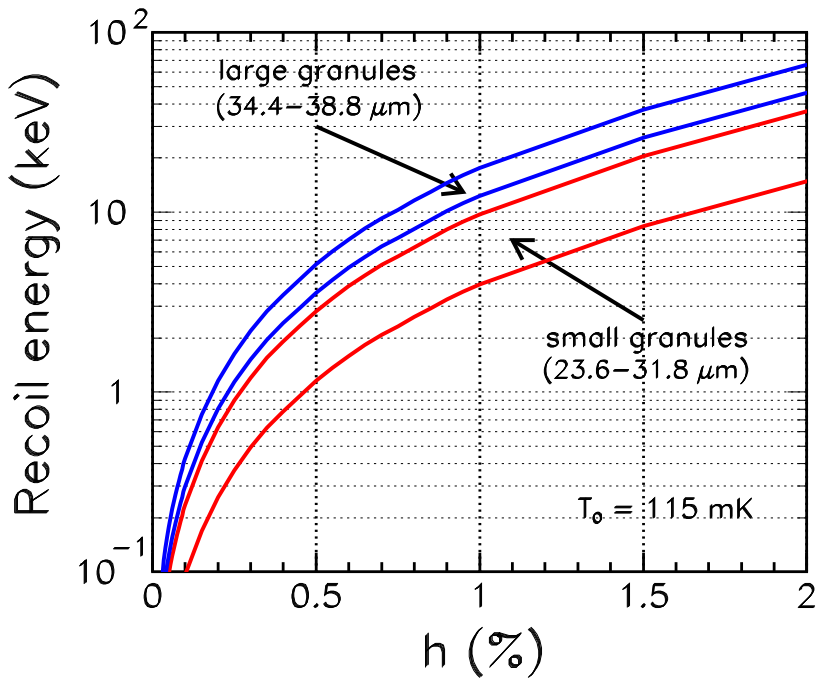}
\caption{Estimated granule energy thresholds, assuming global heating model, 
as a function of magnetic threshold, $h$, for several granule sizes at 115~mK. 
\label{thresh}}
\end{figure}

The detector threshold is set up by ramping the magnetic field up to a 
value $H_2 = 285$~G and then lowering it to an operating point $H_1$, thus 
defining the relative magnetic threshold, $h \equiv 1 - H_1/H_2$. 
Assuming that the entire granule volume needs to be heated (global heating), 
the magnetic threshold corresponds to an effective energy threshold for each
granule size as can be seen in Fig.~\ref{thresh} (see also Eq.~\ref{global}).
The sensitivity of SSG detectors to minimum--ionising particles~\cite{mip}, 
X rays~\cite{xrays}, and $\alpha$ particles~\cite{alpha1,alpha2} has been 
proved in past experiments. The response of SSG devices to nuclear recoil 
energies down to 1~keV has been explored irradiating SSG granules with a 
70~MeV neutron beam \cite{neutron}.

\begin{figure}[h]
\centering
\epsfysize=9.3truecm
\epsffile{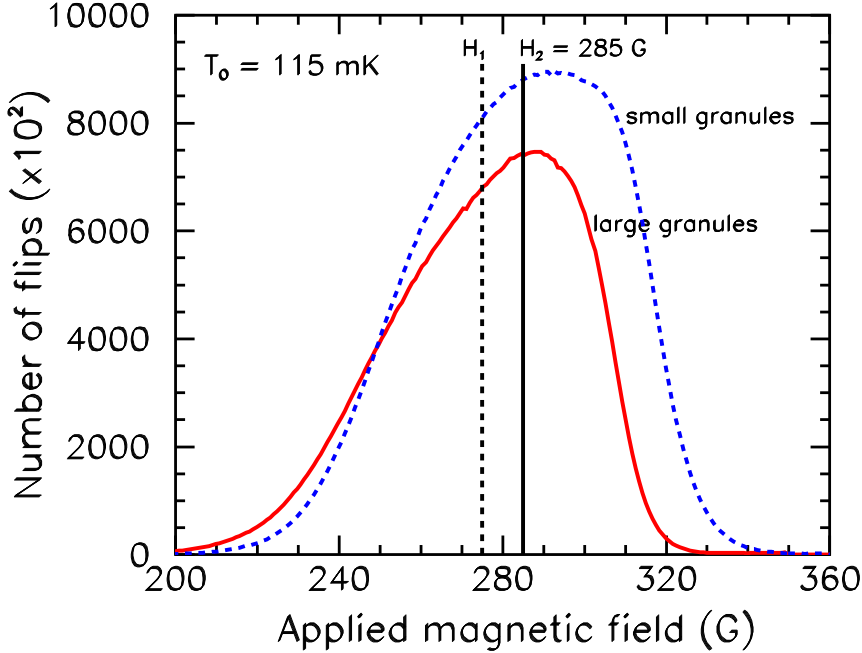}
\caption{Superheating field distributions at $T_0$=115 mK. \label{sh}}
\end{figure}

Unfortunately, not all the granules flip at the same magnetic field, but
instead they exhibit a distribution of values at which the phase transition
occurs, this distribution is the so--called superheating curve. 
This in turn, has the effect of blurring the effective energy threshold of the
detector. A superheating curve characterising our granules, can be measured 
by slowly increasing the magnetic field and recording the number of phase
transition signals as a function of magnetic field as it is shown in
Fig.~\ref{sh}. The superheating spread has a relative width of 22~\%
at FWHM. The figure also shows, schematically, the values defining the 
magnetic threshold.

First results of the 2002 run were published in 
Refs.~\cite{taup01,idm02,ltd10}. In the 2003 run, we measured at magnetic 
thresholds as low as $h =$ 0.075\% and 0.1\%. Much lower threshold values
were difficult to achieve because magnetic field and temperature 
instabilities start to be important.

\subsection{Data analysis}

\begin{figure}[h]
\centering
\epsfysize=10truecm
\epsffile{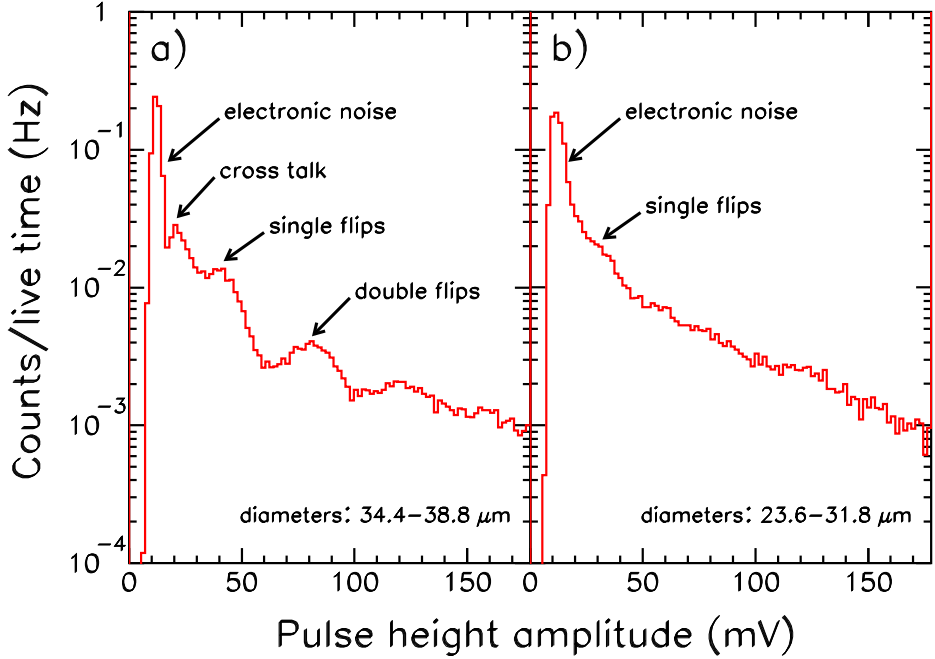}
\caption{Pulse height histograms for large, a), and
small granules,~b). \label{raw}}
\end{figure}

The raw data of a typical run is depicted in Fig.~\ref{raw}, where a histogram
of pulse heights of all digitised pulses in the run is shown. At the 
lowest amplitudes, the signals due to electronic noise dominate the total 
rate of $\approx$ 1~--~10~Hz. For the large granules (Fig.~\ref{raw}a and 
Fig.~\ref{flip}) the signals due to single granule flips in a coil (or 
single flips) are well distinguished from those signals having multiple 
granule flips in the same coil (double flips or even larger). For the small 
granules (Fig.~\ref{raw}b) the difference between single and double flips is 
not so evident due to their wider granule size distribution.

An additional cross--talk noise peak is also seen in Fig.~\ref{raw}a. Noise
pulses are characterised by their shape and low amplitudes, while cross--talk
signals exhibit a polarity opposite to that  of regular flips. 
Cross talk and noisy periods associated with strong vibrations, for instance
during  N$_2$ filling, are easily recognised and not considered in further
analysis.

\subsection{Pulse shape analysis}

\begin{figure}[h]
\centering
\epsfysize=9.4truecm
\epsffile{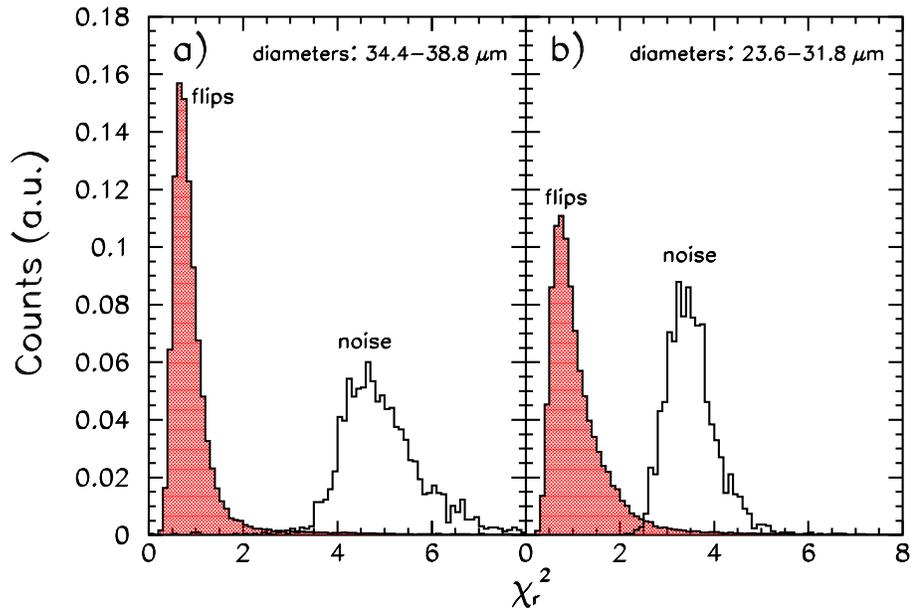}
\caption{Reduced chi--square distributions for arbitrary samples of flip and
noise pulses of large, a), and small granules,~b). Each curve is normalised
to a unit area. \label{chi}}
\end{figure}

\begin{figure}[h]
\centering
\epsfysize=9.4truecm
\epsffile{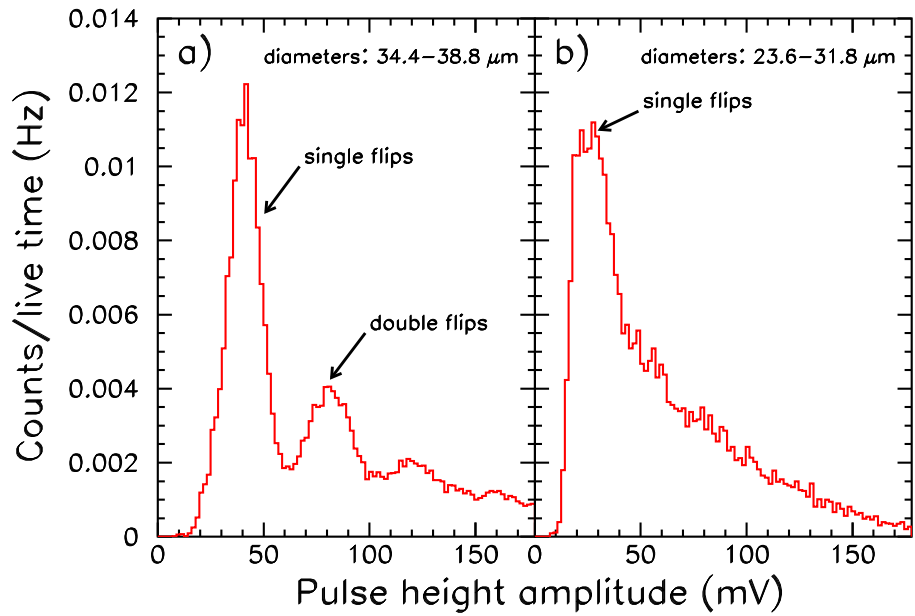}
\caption{Filtered pulse height histograms for large, a), and small
granules,~b). \label{fltrd}}
\end{figure}

To efficiently remove the noise pulses from the data, each individual signal
is compared off--line to an average standard pulse \cite{idm02}. A reliable 
standard pulse is obtained from an average of 500 magnetically induced flips.
For each coil this procedure was repeated, but no differences in the standard
pulses of different channels were found. The digitised signals are fitted to
the standard pulse and only those pulses with a reduced chi square value of
$\chi^2_r \approx 1$ are selected. Figure~\ref{chi} shows the resulting 
chi--square distributions for arbitrary samples of flips and noise signals
for large, a), and small granules, b). We can appreciate from the figure that
a selection criterion of $\chi^2_r< 2.5$ for the large granules and of 
$\chi^2_r< 1.8$ for the small granules still allows a good separation between
the two samples even for the small diameter granules. The original histograms
of Fig.~\ref{raw} are shown again in Fig.~\ref{fltrd}, after the noise and 
cross--talk pulses have been removed and after a $\chi^2_r$ cut has
been applied. Since single flips are not easily distinguished from multiple 
flips, for small granules, a Gaussian fit to the single flip peak was used to 
separate singles from multiples.

~
\subsection{Muon rate}

\begin{figure}[h]
\centering
\epsfysize=9.3truecm
\epsffile{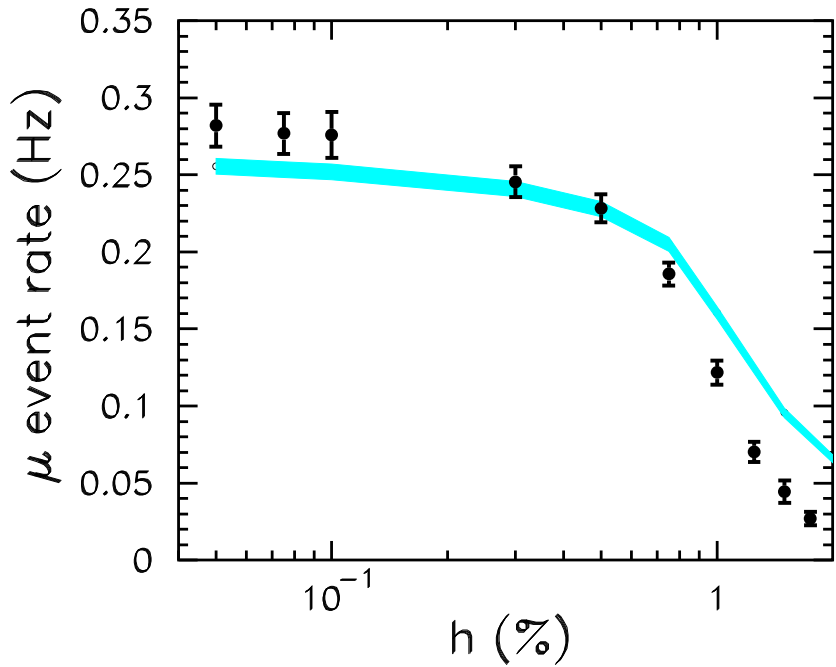}
\caption{Event rate of SSG signals in coincidence with the muon counters
as a function of magnetic threshold, $h$. The shaded region shows the
results of GEANT4 simulations. \label{muons}}
\end{figure}

The SSG signals induced by muons provide a consistency check for the correct 
operation of the detector and for the efficiency of the granules for 
minimum--ionising particles. The event rate of registered SSG signals in
coincidence with the muon counters (accidental coincidences subtracted) is 
shown in Fig.~\ref{muons} as a function of magnetic threshold. Also shown in 
the figure are the results of GEANT4~\cite{geant4} simulations which agree
roughly with the expected response of the detector for minimum ionising
particles.

The muon flux intensity per unit area derived from Fig.~\ref{muons} is 
in good agreement with an independent previous measurement. This measurement, 
was done at the underground site with a simpler and smaller scintillator 
hodoscope arrangement before ORPHEUS was installed.

\section{Results}

\begin{figure}[h]
\centering
\epsfysize=10truecm
\epsffile{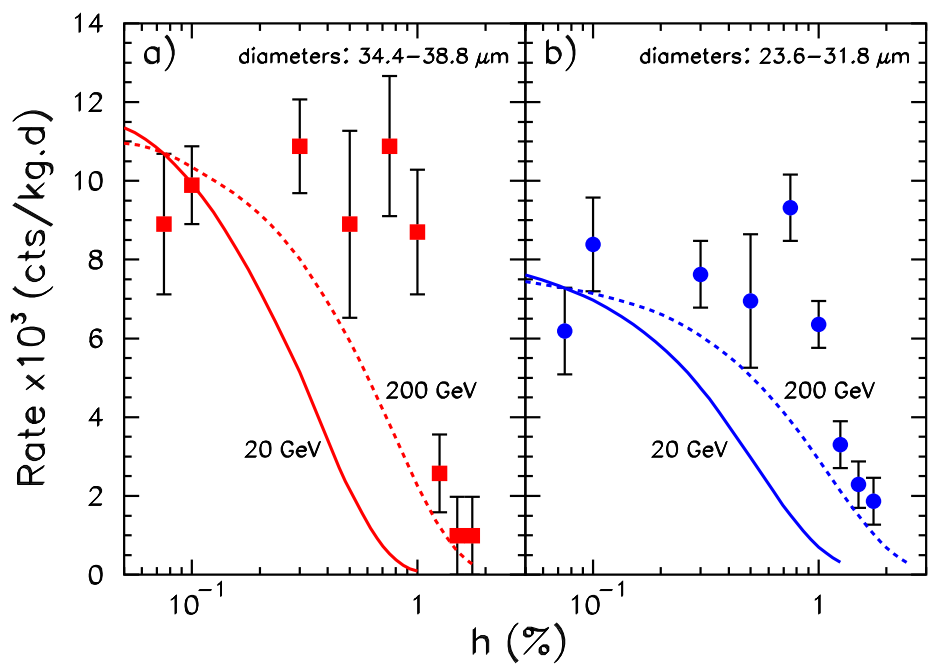}
\caption{Rate of flips with multiplicity $M = 1$ not in coincidence with
muons for large, a), and small granules,~b). Also shown for reference are
the expected WIMP rates for masses of 20~GeV (solid lines) and 200~GeV (dotted
lines). Expected rates are normalised to the upper bounds of the lowest 
thresholds. 
\label{singles}}
\end{figure}

The main result of this experiment relies on efficiently removing 
background events. While gamma rays and through--going muons are
expected to hit more than one granule, ideally, WIMPs are expected
to hit only one single granule. The timing information as well as the
coil location of the recorded pulses in each event are used to asses
the number of flips per event (multiplicity, $M$). Fig.~\ref{singles} 
shows the rate of events having one single flip in only one coil 
($M=1$) not in coincidence with a muon signal, as a function of 
detector threshold. As an example, the figure also shows the expected WIMP
rates (see Appendix) for two neutralino masses: 20~GeV (solid lines) and 
200~GeV (dotted lines). The WIMP cross sections of each curve were chosen 
so as to give expected rates equal to the upper bounds of the most sensitive
points, namely the points with the lowest magnetic thresholds. The error bars
are mostly statistical due to the short measuring times. The total exposure 
for $h = 0.075$\% threshold, in units of kg.d, is 6.69$\times 10^{-3}$.

\vspace{-0.5cm}

Neutrons which produce single flip events are an irreducible source of 
background in this type of experiments. In our underground laboratory, 
neutrons from the rock with energies between 0.5 -- 10~MeV have been 
measured. The measurements yield a flux of ambient neutrons of 
2.4$\times10^{-4}$~n~cm$^{-2}$~s$^{-1}$. This flux turned out to be of the 
same order of magnitude as the neutron flux induced by muons being 
captured in the lead. The 18~cm  thick boron--doped polyethylene shielding
suppresses the total neutron flux by about two orders of magnitude~\cite{natt}.
Thus, the flux reaching the detector is estimated to be 
5$\times10^{-6}$~n~cm$^{-2}$~s$^{-1}$. In terms of neutrons per
kg.d this gives a total neutron background of approximately 
10~n~kg$^{-1}$~d$^{-1}$.

\subsection{Exclusion plots}

\begin{figure}[h]
\centering
\epsfxsize=10truecm
\epsffile{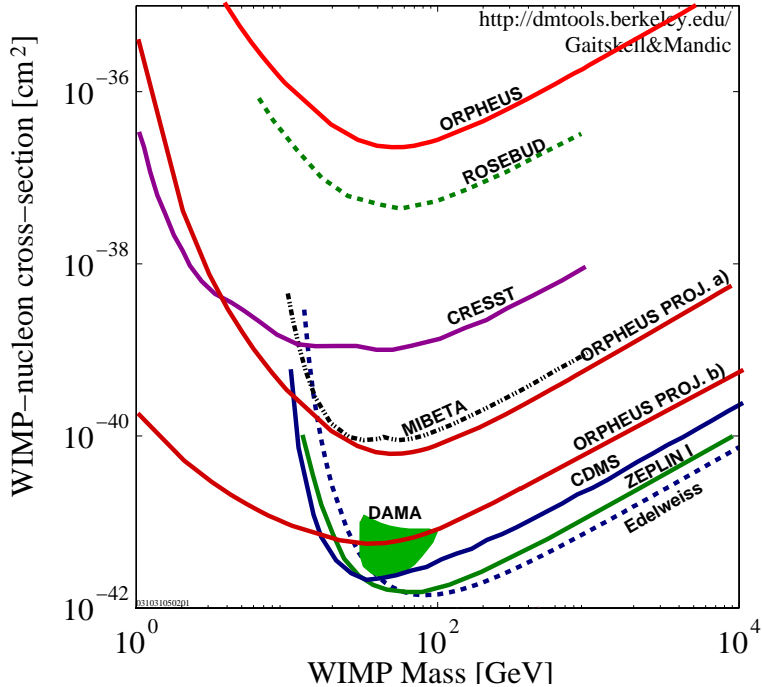}
\caption{ORPHEUS spin--independent exclusion plot (ORPHEUS line)
together with several other experiments 
\cite{EDELWEISS,ZEPLINI,CDMS,MIBETA,CRESST,ROSEBUD,DAMA} (labeled lines).
Projected exclusion plots for ORPHEUS under two different conditions: 
ORPHEUS PROJ. a) background of 1~cnt~kg$^{-1}$~d$^{-1}$ and ORPHEUS PROJ. b) 
1~cnt~kg$^{-1}$~d$^{-1}$, 10~$\mu$m granule diameter, and a superheating 
distribution with FWHM of 1~\%. 
\label{exclus}}
\end{figure}

Comparing the rate of single flips measured with theoretical estimations 
of the expected spin--independent neutralino rate in our detector, an 
exclusion plot, like the one shown in Fig.~\ref{exclus} (ORPHEUS line), 
can be made. The distributions of granule sizes and of superheating fields
are taken into account when calculating the expected neutralino rate (see 
Appendix). Fig.~\ref{exclus} shows the spin--independent neutralino--nucleon
elastic cross section versus neutralino mass assuming a local WIMP halo 
density $\rho_{\chi} = 0.3$~GeV~cm$^{-3}$, an average velocity of the Earth 
in the halo rest frame $V_E = 243.5$~km~s$^{-1}$, and an isothermal spherical 
halo model with a velocity dispersion $\sigma_{v} = 270$~km~s$^{-1}$. The 
region above the curve is excluded at a 90\% confidence level. For our 
exclusion plot, we considered the small granule single rates from 
Fig.~\ref{singles}. Exclusion plots from several other experiments 
\cite{EDELWEISS,ZEPLINI,CDMS,MIBETA,CRESST,ROSEBUD} (labeled lines) are also 
shown in the figure together with the 3$\sigma$ annual modulation region, 
claimed by the DAMA group \cite{DAMA} (grey area). In the constrained minimal
supersymmetric extension of the standard model \cite{Ellis2}, the 
spin--independent neutralino cross sections lie more than 2 orders of 
magnitude below the sensitivity of actual experiments.

\section{Discussion and outlook}

Due to the relatively high background of our experiment the excluded region
in Fig.~\ref{exclus} lies above those of other experiments. In the same figure,
projected exclusion plots are shown under several assumptions: a) same
detector characteristics of the present setup, but a background level 
of 1~cnt~kg$^{-1}$~d$^{-1}$ (attainable at a deeper underground site and with
better shielding), b) 1~cnt~kg$^{-1}$~d$^{-1}$, smaller granules ($\phi = 
10$~$\mu$m) and a narrower relative spread in the superheating distribution 
(FWHM = 1~\%). Narrower superheating distributions enhance the sensitive 
mass of the detector and improve at the same time the background rejection
capability. The use of smaller granules allows smaller energy depositions to 
be detected which makes the detector also more sensitive to smaller WIMP 
masses.

A hint that narrower superheating distribution could be obtained, has been 
shown by us in Ref.~\cite{cyl} (FWHM $\approx$ 6~\%), where a regular array 
of cylinders, produced with an evaporation method, was used. Also, thermal 
treatment of the granules with a laser beam has shown to have some effect in
reducing the spread of the superheating curve \cite{spimp}. Detecting single
flip signals of smaller granules would require more and smaller pickup coils
or a different readout system. The use of superconducting quantum interference
devices (SQUID) could offer such a possibility. In the past, already granules
of 20$\mu$m in diameter were measured in a large size prototype  \cite{squid}.

\section*{Acknowledgements}
We are gratefull to T.~Ebert and K.~U.~Kainer from the Technical University of
Clausthal (Germany) for the production of the tin granules. We would like to 
thank M.~Hess, S.~Lehman, H.~Ruetsch and H.~U.~Sch\"utz for their sustained 
technical support. We are in particular acknowledging the outstanding efforts
in the maintenance and operation of the cryogenics by F.~Nydegger. We also 
would like to thank the Paul Scherrer Institute (Villigen) for providing us 
with liquid He. This work was supported by the Schweizerische Nationalfonds 
zur F\"orderung der wissenschaftlichen Forschung.

\appendix

\section{Expected WIMP rates in SSG detectors}

Tin granules are type I superconductors with a phase diagram of the form
$H(T) \propto [1 - (T/T_c)^2]$ and $T_c = 3.72$~K. A granule initially 
in the superheated superconducting region close enough to its phase transition
boundary, say at ($T_{\rm o}$, $H_{\rm o}$), requires only a small temperature
increase, $\Delta T = T_f - T_{\rm o}$ to become normal. The temperature
increase can be produced by a WIMP interacting with a granule of radius $r_g$
and depositing an energy, 
\begin{equation}
  E_t \;=\; \frac{4 \pi r_g^3}{3} \, \int_{T_{\rm o}}^{T_f} C_V(T) \, dT ,
  \label{global}
\end{equation}
which represents the minimum deposited energy required to induce a flip. At 
low enough temperatures, the specific heat of the superconducting granules is
dominated by its lattice contribution, $ C_V(T) = \beta (T/\theta_D)^3$,
where $\theta_D = 195$~K is the Debye temperature for tin and $\beta = 12 
\pi^4 R /5$, $R = 8.314$~J~mol$^{-1}$~K$^{-1}$. We see therefore that SSG
acts as a threshold detector and that its sensitivity depends on the
granule size, working temperature, and distance to the phase transition 
boundary.

WIMPs of mass $m_\chi$ and velocity $v$ elastically scattered off a nucleus
of mass $m_N$ can impart to it a recoil energy $E = \mu^2 v^2 (1 - 
\cos\theta)/m_N$, where $\mu$ is the reduced mass of the WIMP-nucleus system
and $\theta$ the scattering angle in the centre of mass system. The expected 
WIMP recoil spectrum is given in \cite{Jungman},
\begin{equation}
\frac{dR}{dE} \; = \; \frac{\sigma_{\rm o} \, \rho_{\chi} }{2 \mu^2 \, m_\chi} 
     \; F^2(E) \, \int_{v_{min}}^{v_{max}} \frac{f(v)}{v} \, dv ,
\end{equation}
where $\sigma_{\rm o}$ is the total elastic cross section at zero momentum
transfer, $\rho_{\chi}$ is the local halo density of WIMPs, $F^2(E)$ the 
nuclear form factor, and $f(v)$ the velocity distribution of the WIMPs in the 
halo for an observer on the Earth. $v_{min}$ is the minimum velocity which 
contributes to the recoil energy $E$, namely, $v_{min} = \sqrt{m_N E/ 2 
\mu^2}$, and $v_{max} = 570$~km~s$^{-1}$ the escape velocity from the halo.

\begin{figure}[h]
\centering
\epsfysize=10truecm
\epsffile{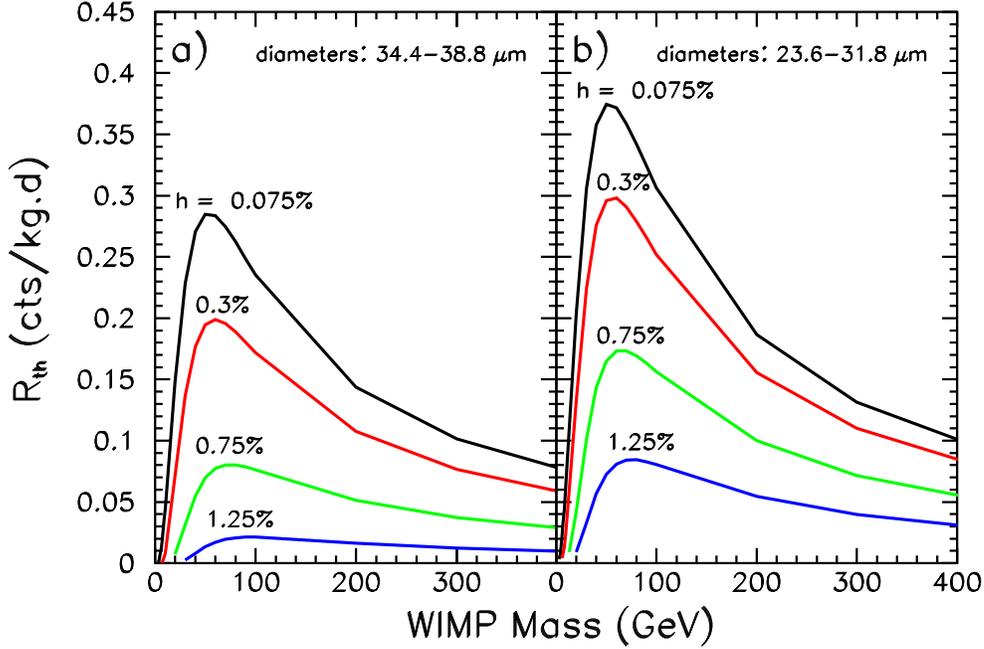}
\caption{Expected WIMP rates as a function of WIMP mass and several detector
thresholds, $h$ for large, a), and small granules,~b). A temperature of 115~mK
and a WIMP--nucleon cross section $\sigma_{\chi p} = 10^{-41}$~cm$^2$ 
have been assumed.
\label{pred}}
\end{figure}

The ORPHEUS detector is made of a collection of granules each being sensitive
to a slightly different energy due to the spread in their superheating fields.
The superheating distribution, $dN_{sh}/dH$, measured experimentally in 
section 3, is therefore equivalent through relation (\ref{global}) to a 
distribution of threshold energies, $dN_{sh}/dE_t$. The expected number of 
flips in our detector, $R$, is thus given by the convolution of the expected 
integral rate with the distribution of threshold energies, which in turn can 
be related to the superheating distribution, 
\begin{equation}
\hspace{-0.3cm}
R_{th} \;=\; \int_0^{E_f} \frac{dN_{sh}}{dE_t} \left( 
	\int_{E_t}^{E_{max}} \frac{dR}{dE}\, dE \right) \, dE_t 
        \;=  \int_0^{E_{max}} \frac{dR}{dE} \left( 
	\int_{H_{\rm o}}^{H(E)} \frac{dN_{sh}}{dH}\, dH \right) \, dE.
\end{equation} 

We parametrise the higher fields part of the superheating distribution of 
Fig.~\ref{sh} using a Gaussian distribution which is simpler to integrate.
We also take into account the fact that the granules have a Gaussian 
distribution of sizes. Considering all this, Fig.~\ref{pred} shows the 
expected rate of WIMP events in our detector for large and small granules,
for different magnetic thresholds $h$, for a given WIMP--nucleon cross section,
$\sigma_{\chi p} = 10^{-41}$~cm$^2$ and as a function of WIMP mass. These 
expected WIMP rates can be directly compared with our experimental results
(see Fig.~\ref{singles}).

\end{document}